\documentclass[%
 aip,
 apl,%
 amsmath,amssymb,
reprint,%
]{revtex4-1}

\usepackage{graphicx}
\usepackage{dcolumn}
\usepackage{bm}
\usepackage{epstopdf}
\usepackage{amsmath}
\usepackage{braket}

\begin{document}

\preprint{AIP/123-QED}

\title[The energy landscape of a superconducting double dot]{The energy landscape of a superconducting double dot}

\author{N. J. Lambert}
\author{M. Edwards}
\author{A. A. Esmail}
\affiliation{Microelectronics Group, Cavendish Laboratory, University of Cambridge, Cambridge, CB3 0HE, UK}
\author{F. A. Pollock}
\affiliation{Atomic \& Laser Physics, Clarendon Laboratory, University of Oxford, Parks Road, Oxford, OX1 3PU, UK}
\author{B. W. Lovett}
\affiliation{SUPA, School of Physics and Astronomy, University of St Andrews, KY16 9SS, UK}
\author{A. J. Ferguson}
\affiliation{Microelectronics Group, Cavendish Laboratory, University of Cambridge, Cambridge, CB3 0HE, UK}
\email{ajf1006@cam.ac.uk}
\date{\today}

\begin{abstract}
We consider the energetics of a superconducting double dot, comprising two superconducting islands coupled in series via a Josephson junction. The periodicity of the stability diagram is governed by the competition between the charging energy and the superconducting gap, and the stability of each charge state depends upon its parity. We also find that, at finite temperatures, thermodynamic considerations have a significant effect on the stability diagram.\end{abstract}

\maketitle

The entangled quasiparticles resulting from split Cooper pairs may form a resource for quantum circuits. Recent theoretical\cite{Recher2001,Lesovik2001,Samuelsson2003,Falci2001,TorresJandMartin1999} and experimental\cite{Hofstetter2009,Herrmann2010,Das2012} studies have described systems in which a supercurrent is divided via two independent quantum dots. In these experiments, Cooper pairs are split due to the charging energies of the quantum dots overcoming the superconducting gap, and the correlation of the noise in the resulting quasiparticles currents is measured. 

Conversely, in superconducting qubits such excitations from the ground Cooper pair state are generally undesirable as they lead to reduced coherence times\cite{Lutchyn2006,Catelani2011}, and may also be relevant to the decoherence of Majorana particles in superconductor/semiconductor hybrids\cite{Rainis2012}. This is termed quasiparticle poisoning. It has been investigated by using transport and charge sensing measurements on superconducting single electron transistors (SSETs)\cite{Knowles2012,Maisi2013} and Cooper pair boxes\cite{Saira2010} to probe the Cooper pair splitting and recombination processes. The competition between the charging energy of the SSET, $E_{C}$, which can act to separate Cooper pairs, and the superconducting gap results in parity effects\cite{Tuominen1992,Lafarge1993a,Lafarge1993,Matveev1993,Joyez1994,Yamamoto2006}; an odd number of charges on the island will result in an unpaired quasiparticle and an additional energy cost of $\Delta$.

In this Letter we describe the energies of quasiparticle and Cooper pair charge states of a superconducting double dot (SDD), which may allow the investigation of both desirable and undesirable quasiparticles. The device consists of two superconducting islands connected by a tunnel barrier, which forms a Josephson junction. The islands are also connected via NIS tunnel barriers to metallic source and drain leads, which act as charge reservoirs and can also be used to probe the impedance of the device. Each island is capacitatively coupled to an electrostatic gate, which may be used to control the chemical potential of the island.
	
\begin{figure}
\includegraphics{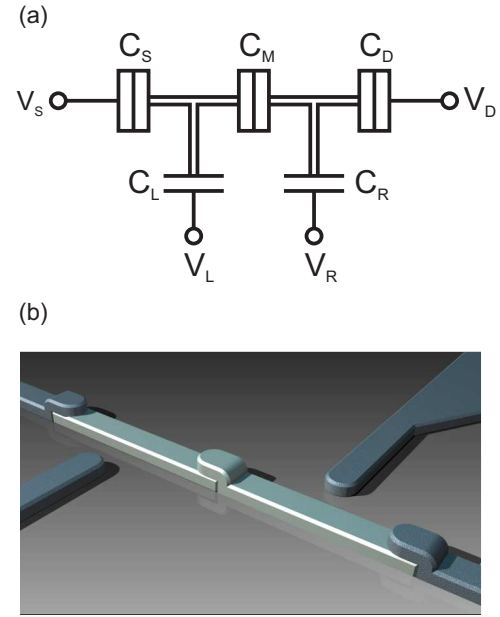}
\caption{The device. (a) The equivalent circuit for the device, with superconducting parts shown as double lines, and normal state parts as single lines. We treat the electrostatic gates and islands coupled by tunnel junctions as a network of capacitances, ignoring cross-capacitances from the left gate to the right island and vice versa. (b) Schematic of a device realised by multiple angle evaporation. Superconducting islands are shown in light grey, and normal state gates and leads in dark grey.}
\label{fig1}
\end{figure}

The equivalent circuit for the device is shown in Fig.\ \ref{fig1}, and we label the gate ($C_L$, $C_R$), lead ($C_S$, $C_D$) and inter-island ($C_M$) capacitances. We neglect any cross capacitances which couple the right lead to the left dot and vice versa.	Such a device can be readily realised by electron beam lithography followed by multiple-angle evaporation and controlled oxidation of aluminium\cite{Dolan1977,Lambert2013}.

The behaviour of the device is governed by the competition between six energy scales: the superconducting gap ($\Delta$), the charging energies of the islands\cite{VanDerWiel2002} ($E_{CR}$, $E_{CL}$,$E_{CM}$), the temperature ($k_BT$) and the Josephson energy of the middle tunnel junction ($E_J$). Following the approaches of Tuominen\cite{Tuominen1992} and Lafarge\cite{Lafarge1993a,Lafarge1993} we now calculate the Helmholtz free energy for different charge states of the device, $F=U-TS$, where $U$ is the internal energy of the system, $T$ is the temperature and $S$ is the entropy.

We label the charge states $(m,n)$ where $m$ ($n$) is the total offset charge from an arbitrarily chosen \textit{even parity} state on the left (right) island. An important distinction exists between odd and even parity states. In even parity states the ground state is a Cooper pair condensate and is separated from other states by $2\Delta$. The state can be unambiguously labelled $(m,n)$, where $m$ and $n$ have even parity. If, however, either $m$ or $n$ is odd, then the associated quasiparticle has a momentum degree of freedom. We neglect the quantisation of quasiparticle states due to lateral confinement; typical island lengths in aluminium SETs are $\sim1$ $\mu$m, and so we estimate the energy separation of the lowest momentum states to be around $\frac{3 \pi ^2 \hbar ^2}{2 L^2 m}\approx 1$ $\mu$eV. Therefore the quasiparticle states form a continuous band. We initially concentrate on the lowest-lying quasiparticle momentum state.

To determine the internal energy of the device we now solve the Hamiltonian. Diagonal elements of the Hamiltonian are given by the sum of the electrostatic energy, and the energy cost of any unpaired quasiparticles. The electrostatic energy is found by treating the device as a network of capacitances\cite{VanDerWiel2002}, and is given by $U_E(m,n)=\frac{1}{2} \mathbf{V} \cdot \mathbf{Q}$, where $\mathbf{V}$  and $\mathbf{Q}$ are vectors, the components of which are the voltages and charges on each of the nodes. The internal energy is then given by

\begin{align*}
U(m,n)=U_E(m,n) + \Delta(m \bmod{2}  + n \bmod{2})
\end{align*}

Off-diagonal elements are also dependent on the parity of the charge states involved. If the charge states are coupled by quasiparticle tunnelling, then the relevant matrix element is zero because the quasiparticles can occupy a continuum of degenerate momentum states. If, on the other hand, the states are separated by the transfer of one Cooper pair between the islands, then the matrix element is $-E_J/2$\cite{Likharev1985,Duty2005}.

We can now write the Hamiltonian in matrix form for a subspace spanning the nine states with $n = 0\rightarrow2$, $m = 0\rightarrow2$.

\begin{widetext}
\begin{align*}
\hat{H}=\left( \begin{array}{ccccccccc}
U_E(0,0) & 0 & 0 & 0 & 0 & 0 & 0 & 0 & 0 \\
0 & U_E(1,0)+\Delta & 0 & 0 & 0 & 0 & 0 & 0 & 0 \\
0 & 0 & U_E(2,0) & 0 & 0 & 0 & -E_J/2 & 0 & 0 \\
0 & 0 & 0 & U_E(0,1)+\Delta & 0 & 0 & 0 & 0 & 0 \\
0 & 0 & 0 & 0 & U_E(1,1)+2\Delta & 0 & 0 & 0 & 0 \\
0 & 0 & 0 & 0 & 0 & U_E(2,1)+\Delta & 0 & 0 & 0 \\
0 & 0 & -E_J/2 & 0 & 0 & 0 & U_E(0,2) & 0 & 0 \\
0 & 0 & 0 & 0 & 0 & 0 & 0 & U_E(1,2)+\Delta & 0 \\
0 & 0 & 0 & 0 & 0 & 0 & 0 & 0 & U_E(2,2)
\end{array}
\right)
\end{align*}
\end{widetext}
	
The eigenvalues of this Hamiltonian are $U(m,n)$, except for the single Cooper pair subspace. The eigenvalues for these have the form

\begin{align*}
 U =\ &\frac{1}{2} \Bigl(U_E(2,0)+U_E(0,2)\ \pm\\
 &\sqrt{U_E(2,0)^2-2 U_E(2,0) U_E(0,2)+U_E(0,2)^2+E_J^2}\Bigr).
\end{align*}

This concludes our calculation of the internal energy of the different charge states.

We now determine the contribution of the entropy to the free energy, following the approach of Tuominen \textit{et al}\cite{Tuominen1992}. The entropy is given by $S=k_BT\ln(N_{\textrm{eff}})$, where $N_{\textrm{eff}}$ is the effective number of microstates available to the system. For Cooper pair charge states there is only one available state and the entropic contribution to the free energy is zero. Quasiparticle states, on the other hand, occupy a continous band of momentum states. In the case where a quasiparticle is present on an island, the number of microstates available is

\begin{align*} 
N_{\textrm{eff}}= 2 V D(\epsilon_{F})\int^{\infty}_{\Delta}f(\epsilon)g(\epsilon) d\epsilon
\end{align*}

where $V$ is the island volume, $D$ is the metallic density of states, $\epsilon_{F}$ is the energy at the Fermi surface, $f(\epsilon)$ is the Fermi distribution and $g(\epsilon)$ is the BCS density of states. At low temperatures, the total number of microstates for each quasiparticle becomes $N_{\textrm{eff}}\approx2\sqrt{2}VD(\epsilon_{F})\sqrt{\Delta k_{B}T}$.

The entropy is therefore also parity dependent, and we can now write the free energy of the system as

\begin{align*}
F(m,n)=U_{E}(m,n)+\tilde{\Delta}(m \bmod{2}  + n \bmod{2})
\end{align*}

where $\tilde{\Delta}=\Delta-k_{B}T\ln (N_{\textrm{eff}})$. The energy of quasiparticle states is therefore suppressed as temperature is increased, with the energy of two-quasiparticle states decreasing at twice the rate of single-quasiparticle states.

\begin{figure}
\includegraphics{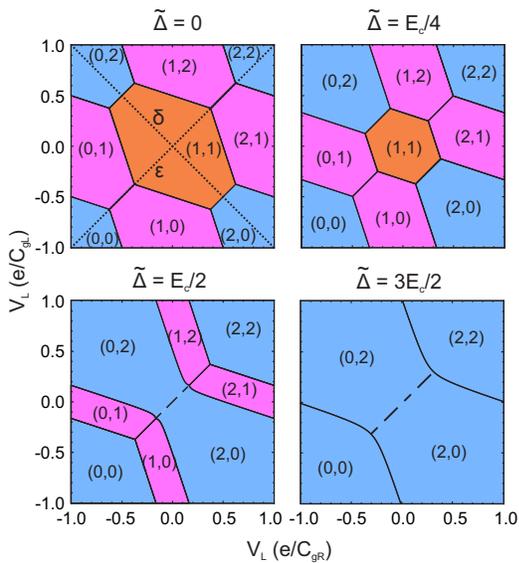}
\caption{Charge stability diagrams for (a) $\tilde{\Delta}=0$ (b) $\tilde{\Delta}=E_C/4$ (c) $\tilde{\Delta}=E_C/2$ (d)$\tilde{\Delta}=3E_C/2$. Blue regions are Cooper pair states, purple regions have one quasiparticle present and the orange region has one quasiparticle on each island. Dotted lines in the top left panel show the $\delta$ and $\epsilon$ axes.}
\label{fig2}
\end{figure}

\begin{figure}
\includegraphics{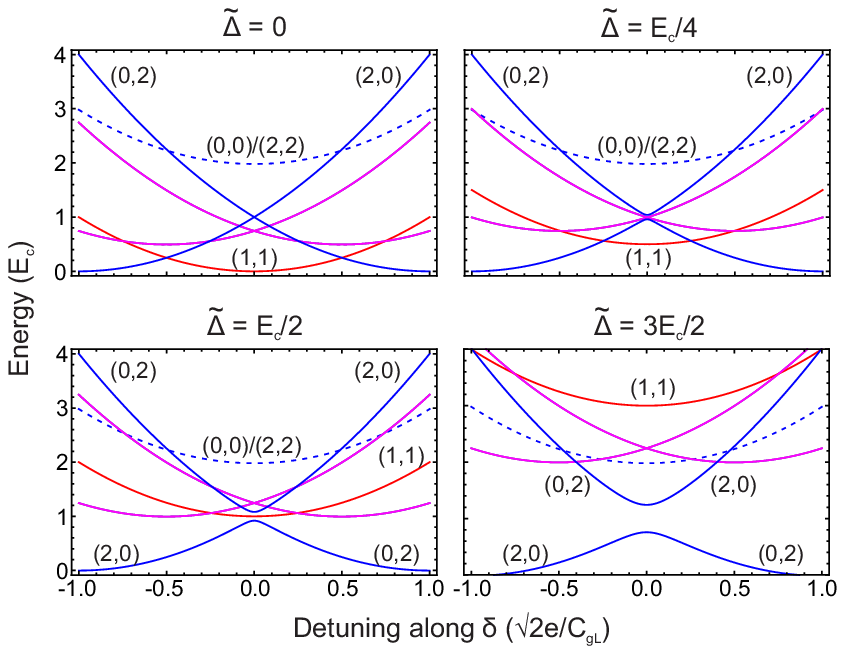}
\caption{Cross-sections along $\delta$ in Fig.\ 2 for (a) $\tilde{\Delta}=0$ (b) $\tilde{\Delta}=E_C/4$ (c) $\tilde{\Delta}=E_C/2$ (d)$\tilde{\Delta}=3E_C/2$. Colours are as for Fig. 2.}
\label{fig3}
\end{figure}

\begin{figure}
\includegraphics{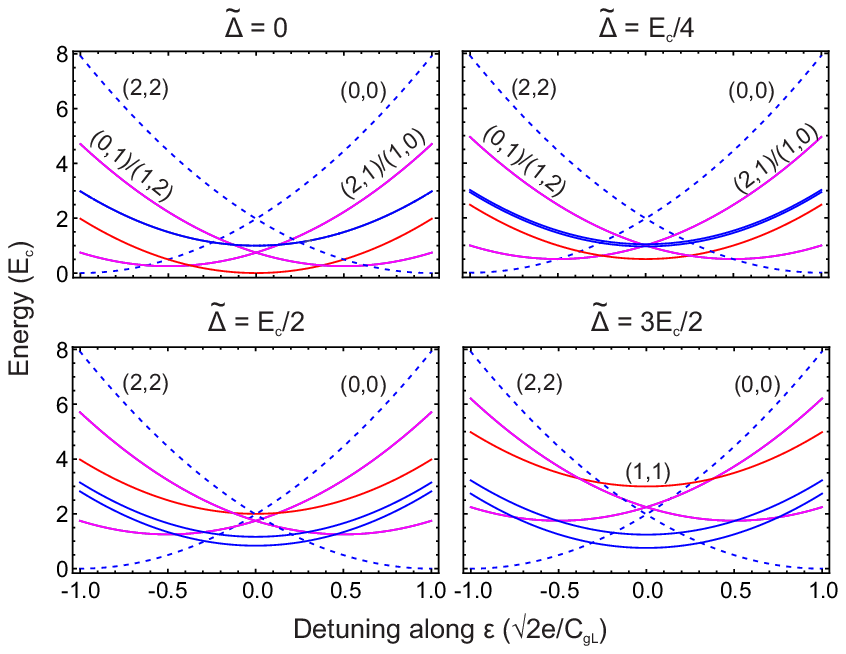}
\caption{Cross-sections along $\epsilon$ in Fig.\ 2 for (a) $\tilde{\Delta}=0$ (b) $\tilde{\Delta}=E_C/4$ (c) $\tilde{\Delta}=E_C/2$ (d)$\tilde{\Delta}=3E_C/2$. Colours are as for Fig. 2.}
\label{fig4}
\end{figure}

We show the evolution of the calculated charge stability diagram (Fig.\ \ref{fig2}) and band structure along lines $\delta$ (Fig.\ \ref{fig3}) and $\epsilon$ (Fig.\ \ref{fig4}) as a function of $\tilde{\Delta}/Ec$. We use the Ambegaokar-Baratoff relationship between $E_J$ and $\Delta$, 

\begin{align*}
\label{AmbBara}
E_{J}=\frac{\Delta h}{8 e^2 R_m}
\end{align*}

These cross sections lie along the boundaries between some charge states in the stability diagram and so some pairs of states are degenerate along these lines. Along $\delta$, the degenerate pairs are $\left\{(0,1),(1,0)\right\}$ and $\left\{(1,2),(2,1)\right\}$, and along $\epsilon$ the pairs $\left\{(1,0),(2,1)\right\}$, $\left\{(0,1),(1,2)\right\}$ and $\left\{(0,0),(2,2)\right\}$ are degenerate.

For $\tilde{\Delta}/E_c=0$ the stability diagram is the characteristic $e$ periodic `honeycomb' pattern of a metallic double dot (Fig.\ \ref{fig2}(a)). Increasing $\tilde{\Delta}/Ec$ has two effects: the energy cost for quasiparticles increases, leading to cells with odd parity on either of the islands reducing in size (Fig.\ \ref{fig2}(b)); and an anticrossing of size $E_J$ opens up between states coupled by Cooper pair tunelling, lifting the degeneracy that lies along line $\delta$ (Fig.\ \ref{fig4}(b)).

When $\tilde{\Delta}/(E_c-E_J) > \frac{1}{2}$  then the energy cost of the (1,1) state is high enough that it is never the ground state(Figs. \ref{fig2}(c), \ref{fig3}(c) and \ref{fig4}(c)). Instead, the ground state at ($V_L=0$,$V_R=0$) becomes the symmetric combination of the Cooper pair states, $\Psi=\frac{1}{\sqrt{2}}(\ket{2,0}+\ket{0,2})$.

As $\tilde{\Delta}/E_c$ is increased still further, single quasiparticle states are also expelled from the stability diagram, and it becomes $2e$ periodic. At ($V_L=0$,$V_R=0$), the lowest two states are the symmetric and antisymmetric Cooper pair states which form a two-level system with energy separation $E_J$.

The energy landscape can therefore be modified by adjusting either the temperature or the magnetic field. To illustrate the size of this effect, we now set device parameters typical of superconducting aluminium devices\cite{Ferguson2008,Ferguson2006b,Pekola2010}. We choose $C_S=C_D=0.8$ fF, and $C_M = 0.4$ fF. These are much larger than the gate capacitances ($C_{gL}=C_{gR}\sim1$ aF) and so dominate the charging energies, which are $E_{CL}=E_{CR}=E_C=150$ $\mu$eV and $E_{CM}=50$ $\mu$eV. We choose the superconducting gap at $T=0$, $B=0$ to be $\Delta_0=225$ $\mu$eV $=\frac{3}{2} E_C$. We also set the critical temperature $T_c = 1.2$ K, and $E_F=11.7$ eV. The resistance of the middle tunnel junction is $20$ k$\Omega$, giving a zero field $E_J$ of $24$ $\mu$eV, and we use island dimensions $1$ $\mu\textrm{m}$ $\times$ 100 nm $\times$ 20 nm, giving a volume of $2\times10^{-3}$ $\mu\textrm{m}^3$.

\begin{figure}
\includegraphics{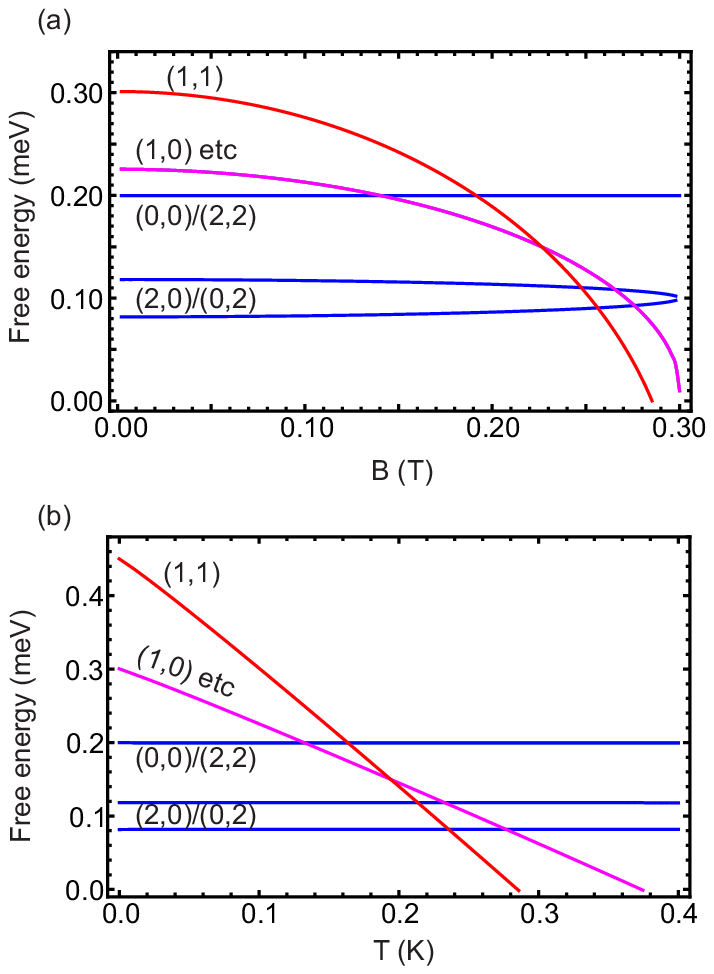}
\caption{Free energies at zero detuning as a function of (a) applied field and (b) temperature. Both increasing field and temperature suppress the free energy of quasiparticle states, changing the ordering of the charge states.}
\label{fig5}
\end{figure}

In Fig.\ 5 we show the energies of different charge states at zero detuning - the middle point of Fig.\ 2. In Fig.\ \ref{fig5}(a) these levels are shown as a function of magnetic field with $T = 100$ mK, and in Fig.\ \ref{fig5}(b) the evolution with temperature at $B=0$ is shown. In both cases, the energies of the quasiparticle states are suppressed. In particular, the ordering of the states changes, with the (1,1) state becoming the ground state as the quasiparticle energy cost decreases. We note for clarity that the reduction in the free energy of quasiparticle states with temperature is not due to thermal supression of $\Delta$, but rather the increase in the entropy factor due to the increase of $N_{\mathrm{eff}}$.

The energies of the charge states at zero detuning can also be written as a function of the charging energies and $\tilde{\Delta}$. In particular, we find that $F(1,1)=2\tilde{\Delta}$ and $F((2,0)\pm(0,2))=\frac{1}{2}(E_{CL}+E_{CR})-E_{CM}\pm E_J$.
	
We have described the energetics of a superconducting double dot, taking into account the electrostatic energy, the contribution from unpaired quasiparticles and the entropic contribution to the free energy. We find that the cells in the charge stability diagram show parity dependent behaviour as $\tilde{\Delta}/E_c$ is changed, and describe the level structure as a function of applied magnetic field and temperature. We anticipate novel experiments that enable the controlled splitting and reconstruction of single Cooper pairs based on this energy landscape.
	
A.J.F. would like to acknowledge the Hitachi Research fellowship, support from Hitachi Cambridge Laboratory and support from the EPSRC grant EP/H016872/1. B.W.L. is supported by a Royal Society University Research Fellowship. F.A.P. would like to thank the Leverhulme Trust for financial support.

\bibliography{DoubleSCenergy}

\end{document}